\documentclass[preprint,aps,amssymb,showpacs]{revtex4}

\usepackage{bm}
\usepackage{subfigure}

\begin{document}

\title{Angular Momentum on the Lattice: The Case of Non-Zero Linear Momentum}

\author{David C. Moore}
\author{George T. Fleming}

\affiliation{Sloane Physics Laboratory, Yale University, New Haven,
CT 06520, USA}

\date{\today}

\begin{abstract}
The irreducible representations (IRs) of the double cover of the
Euclidean group with parity in three dimensions are subduced to the
corresponding cubic space group.  The reduction of these
representations gives the mapping of continuum angular momentum
states to the lattice in the case of non-zero linear momentum. The
continuous states correspond to lattice states with the same
momentum and continuum rotational quantum numbers decompose into
those of the IRs of the little group of the momentum vector on the
lattice. The inverse mapping indicates degeneracies that will appear
between levels of different lattice IRs in the continuum limit,
recovering the continuum angular momentum multiplets. An example of
this inverse mapping is given for the case of the ``moving''
isotropic harmonic oscillator.
\end{abstract}


\maketitle

\section{Introduction}
A question of great interest in lattice QCD is how rotational states
on the lattice correspond to states of definite angular momentum in
the continuum limit. To find this correspondence, we find the
mapping of continuum states to lattice irreducible representations
(IRs) and invert it. This problem has been discussed in several
contexts previously -- in solid state physics, the ``cubic
harmonics" are formed by a projection of the continuum spherical
harmonics onto the lattice, \textit{e.g.}\  \cite{Altmann65}.  In
lattice QCD, the reduction of continuum states to the hypercubic
lattice was given by Mandula \textit{et al.}\
\cite{Mandula83,Mandula83wb}, and the reduction of the full
continuum symmetry group (including Poincar\'{e}, color, flavor, and
baryon number symmetries) to the lattice for the case of staggered
fermions was given by Golterman and Smit
\cite{Golterman84a,Golterman84b} and expanded to include non-zero
momentum by Kilcup and Sharpe \cite{Kilcup86}.

We are interested in the classes of lattice actions with unbroken
flavor symmetries (\textit{i.e.}\  Wilson and overlap).  In
particular, we focus on the IRs of the symmetry group of the lattice
Hamiltonian. Johnson \cite{Johnson82} considered the mapping of
continuum SU(2) states to the octahedral group and its double cover,
and Basak \textit{et al.}\  \cite{Basak05,Basak05b} considered the
inclusion of parity in these groups. We expand on this work to
include non-zero momentum.  This extension will be needed to apply
group theory methods to the calculation of matrix elements for hadrons
and electromagnetic currents on the lattice \cite{Renner02}.

The continuum spacetime symmetries of the QCD action are given by
the Poincar\'{e} group, $\mathcal{P}$.  The spatial symmetries of
the QCD Hamiltonian correspond to a subgroup of $\mathcal{P}$ which
is the semidirect product of the group of orthogonal transformations
in three dimensions and the group of translations, $\mathcal{T}^3
\rtimes \mathrm{O}(3)$ (pure time inversion is also a symmetry
element, but its addition is trivial since it commutes with all
other group elements). On the lattice, the symmetry group is
$\mathcal{T}^3_{lat}\rtimes \mathrm{O}_h$, where the rotational
group is reduced to a subgroup of O(3) with only a finite number of
rotations and reflections, and where the subgroup of $\mathcal{T}^3$
contains only lattice translations. Throughout this paper we
consider the double covers of the rotational groups since the double
valued IRs correspond to fermionic states in the continuum
\cite{Johnson82}. We denote the continuous group $\mathcal{G} =
\mathcal{T}^3 \rtimes \mathrm{O}^\mathrm{D}(3)$ and the discrete
group $\mathcal{T}^3_{lat}\rtimes \mathrm{O}^\mathrm{D}_h$, where
the superscript D denotes the double cover.

We construct the IRs of these groups from the normal subgroups of
translations by the method of little groups. We then subduce a
representation of $\mathcal{T}^3_{lat}\rtimes
\mathrm{O}^\mathrm{D}_h$ by restricting the IRs of $\mathcal{G}$ to
the subgroup of lattice elements.  Using the orthogonality
properties of characters, we reduce the subduced representations
into the IRs of the lattice group.  As in the zero momentum case,
many continuous states are mapped to each angular momentum state on
the lattice. This presents complications in finding the inverse
mapping, which is our actual interest. The ambiguities associated
with this inverse mapping are demonstrated for the simple case of
the moving harmonic oscillator.

\section{The symmetry groups and their representations}
\subsection{The discrete group, $\mathcal{T}^3_{lat}\rtimes \mathrm{O}^\mathrm{D}_h$}
The group of proper rotations of a cube in three dimensions is the
octahedral group, O. We are also interested in its double cover,
$\mathrm{O^D}$, and we consider the inclusion of parity by forming
the group $\mathrm{O}^\mathrm{D}_h = \mathrm{O^D} \times
\mathrm{C_2}$, where $\mathrm{C_2}$ consists of the identity and a
parity element $I_s$, corresponding to inversion of all three
coordinate axes through the origin. The inclusion of parity is
straightforward because $I_s$ commutes with all proper rotations in
three dimensions. Throughout this paper we adopt the Mulliken
convention for the labeling of finite groups.

For $\mathcal{T}^3_{lat}\rtimes \mathrm{O}^\mathrm{D}_h$, we can
then write the group elements $\{R_i, \mathbf{n}\}$, where $R_i$
denotes a rotation in one of the lattice rotation groups discussed
above, followed by a lattice translation by $\mathbf{n}$.  The
subgroup of translations, $\mathcal{T}^3_{lat}$, is normal, so we
can easily use this subgroup to induce the IRs of the full group
\cite{Altmann65,Bouckaert36}. The IRs of $\mathcal{T}^3_{lat}$ are
labeled by a vector of real numbers, $\mathbf{k}$, where the $k_i$
lie in an interval of length $2\pi$.  For an IR of
$\mathcal{T}^3_{lat}$ labeled by $\mathbf{k}$, we form the little
group, which consists of the $R_j$ in $\mathrm{O}^\mathrm{D}_h$ such
that $R_j \mathbf{k} = \mathbf{k}$.  We also form the star of
$\mathbf{k}$, which is the set of vectors $\mathbf{k}_p = R_p
\mathbf{k}$ for all $R_p$ in the rotation group.  The IRs of
$\mathcal{T}^3_{lat}\rtimes \mathrm{O}^\mathrm{D}_h$ are then
labeled by $\mathbf{k}$ and a label $\alpha$ denoting an irreducible
representation of the little group of $\mathbf{k}$. After summing
over the diagonal elements, we find the characters:
\begin{equation}
\label{eq:dischar} \chi^{(\mathbf{k}, \alpha)}(R_j,\mathbf{n}) =
\sum_{\mathbf{k}_q} \chi^{\alpha}(R_j) \mathrm{e}^{-i \mathbf{k}_q
\cdot \mathbf{n}}
\end{equation}
where the sum is taken over only the subset $\mathbf{k}_q$ of the
star vectors $\mathbf{k}_p$ such that $R_j$ is in the little group
of $\mathbf{k}_q$.  Also, $\chi^{\alpha}(R_j)$ is the character in
the representation $\alpha$ of the little group and in general can
depend on the particular point in the star. We see that
representations labeled by $\mathbf{k}$ in the same star are
equivalent, so we label the inequivalent irreducible
representations, $\Gamma^{(|\mathbf{k}|,\alpha)}$, by a star given
by $|\mathbf{k}|$ and an IR of the little group labeled by $\alpha$.
Care must be taken to distinguish between IRs labeled by
$\mathbf{k}$ in different stars but with the same length,
\textit{e.g.}\ $\mathbf{k}=(2,2,1)$ and $\mathbf{k}=(3,0,0)$.

To use the above formula, we will need to know the characters of the
little group for a given $\mathbf{k}$. If $\mathbf{k} = (0,0,0)$,
then the little group is the full rotation group,
$\mathrm{O}^\mathrm{D}_h$, and the results for this case are
well-known. If $\mathbf{k}$ is non-zero, then the little group
becomes a subgroup of $\mathrm{O}^\mathrm{D}_h$. The possible little
groups for the rotation group $\mathrm{O}^\mathrm{D}_h$ are given in
Tab.~\ref{tab:lg}. Since we are interested in the covering group
$\mathrm{SU}(2) \times \mathrm{Z}_2$, we find that the double covers
of the little groups are dicyclic groups \cite{jahn:privcomm}. The
character tables for the double covers of the little groups can be
calculated using the $J^P = \frac{1}{2}^-$ representation of
$\mathrm{SU}(2) \times \mathrm{Z}_2$ to determine the multiplication
tables \cite{Lax74}.

\begin{table}
\caption{\label{tab:lg} Little groups for the possible lattice
momenta. $\mathrm{Dic}_n$ denotes the dicyclic group of order $4n$.}
\begin{ruledtabular}
\begin{tabular}{lcc}

  $(0,0,0)$ & $\mathrm{O}_h$ & $\mathrm{O}_h^\mathrm{D}$ \\
  $(n,0,0)$ & $\mathrm{C_{4v}}$ & $\mathrm{Dic}_4$ \\
  $(n,n,0)$ & $\mathrm{C_{2v}}$ & $\mathrm{Dic}_2$ \\
  $(n,n,n)$ & $\mathrm{C_{3v}}$ & $\mathrm{Dic}_3$ \\
  $(n,m,0)$ & $\mathrm{C_{2}}$ & $\mathrm{C}_4$ \\
  $(n,n,m)$ & $\mathrm{C_{2}}$ & $\mathrm{C}_4$ \\
  $(n,m,p)$ & $\mathrm{C_{1}}$ & $\mathrm{C}_2$ \\

\end{tabular}
\end{ruledtabular}
\end{table}

\subsection{The continuous group, $\mathcal{G}$}
We can construct the characters for the irreducible representations
of the continuous group $\mathcal{G}$ analogously.  Translations are
again a normal subgroup and their IRs are labeled by $\mathbf{k}$,
but there is no longer a restriction on the values of the $k_i$.
Every non-zero vector $\mathbf{k}$ now has a little group of the
same form, which consists of rotations in the plane orthogonal to
$\mathbf{k}$ and reflections through planes containing $\mathbf{k}$.
This little group, known as $\mathrm{C_{\infty v}}$, is isomorphic
to the group O(2) of all orthogonal matrices in two dimensions (as
usual we consider the double covers). We induce a representation of
$\mathcal{G}$ from the subgroup of translations, and we construct
basis vectors labeled by the star of $\mathbf{k}$, and by $m_j =
0^+,0^-,\frac{1}{2},1,...$, which labels the IR of the little group
($\mathrm{C_{\infty v}}$ has two one dimensional IRs labeled as $m_j
= 0^+,0^-$.  All other IRs are two dimensional and are labeled by a
positive integer or half-integer). Here, the star of $\mathbf{k}$ is
all vectors of length $|\mathbf{k}|$. A general irreducible
representation of $\mathcal{G}$ is then labeled as
$E^{(|\mathbf{k}|,m_j)}$. Physically, $m_j$ denotes the projection
of the total angular momentum, $j$, along $\mathbf{k}$.

We are interested in subducing a representation of the cubic space
group by considering the IRs of $\mathcal{G}$ restricted to the
subgroup of elements which are in the lattice group. As before, we
are only interested in the characters of the (generally reducible)
representations of the space group subduced from $\mathcal{G}$. For
these elements, we compute the characters by summing only over the
diagonal matrix elements in the IRs of $\mathcal{G}$ which
correspond to lattice vectors $\mathbf{k}$.  The character of this
subduced representation for a general group element $\{R_i,
\mathbf{n}\}$ is then given as:
\begin{equation}
\label{eq:contchar} \chi^{(\mathbf{k}, \alpha)}(R_i,\mathbf{n}) =
\chi^{(m_j)}(R_i) \sum_{\mathbf{k}_q}\mathrm{e}^{-i \mathbf{k}_q
\cdot \mathbf{n}}
\end{equation}
where, as in the discrete case, the sum is only taken over the
vectors $\mathbf{k}_q$ of length $|\mathbf{k}|$ on the lattice such
that $R_i$ leaves $\mathbf{k}_q$ invariant.  The characters
$\chi^{(m_j)}$ of a rotation $R_i$ of magnitude $\theta$ are given
in Tab.~\ref{tab:cinfvchar}.  In the special case $\mathbf{k} = 0$,
the star contains only one point, and the little group is just
$\mathrm{O}^{\mathrm{D}}$(3). Therefore the representations
$E^{(0,m_j)} = D^{(j,\pi)}$, which are just the familiar
representations of $\mathrm{SU(2)} \times \mathrm{C_2}$ labeled by
$j=0,\frac{1}{2},1,...$ and a parity $\pi = \pm1$.
\begin{table}
\caption{\label{tab:cinfvchar}The character table for the double
cover of $\mathrm{C}_{\infty \mathrm{v}}$. $E$ denotes the identity,
$R(\theta)$ is any proper rotation by $\theta$, and $\sigma_v$ is
any reflection through the axis given by $\mathbf{k}$. }
\begin{ruledtabular}
\begin{tabular}{cccc}
  $m_j$ & $E$ & $R(\theta)$ & $\sigma_v$ \\
  \hline
  $0^+$ & 1 & 1 & 1 \\
  $0^-$ & 1 & 1 & -1 \\
  $m_j\geq\frac{1}{2}$ & 2 & $2\cos m_j\theta$ & 0 \\
\end{tabular}
\end{ruledtabular}
\end{table}

\section{The reduction of states}
By the orthogonality properties of characters for finite groups, the
multiplicity, $m$, that an irreducible representation of
$\mathcal{T}^3_{lat}\rtimes\mathrm{O}^\mathrm{D}_h  $ labeled by
$(|\mathbf{k}|,\alpha)$ is contained in a subduced representation of
$\mathcal{G}$ labeled by $(|\mathbf{k}'|,m_j)$ is given by:
\begin{equation}
\label{eq:proj} m = \frac{1}{g} \sum_a
\chi^{(|\mathbf{k}|,\alpha)}(G_a)^\ast
\chi^{(|\mathbf{k}'|,m_j)}(G_a)
\end{equation}
where the sum is taken over all group elements $G_a$, and $g$ is the
order of the group.  The groups $\mathcal{G}$ and
$\mathcal{T}^3_{lat}\rtimes \mathrm{O}^\mathrm{D}_h$ allow
arbitrarily large translations, so we cannot apply
Eq.~(\ref{eq:proj}) directly, but must consider the 3-torus formed
by the boundary conditions $\mathbf{r} + \mathbf{N} = \mathbf{r}$
for all vectors $\mathbf{r}$ and some constant vector $\mathbf{N} =
(N,N,N)$. The boundary conditions limit the allowed values of
$\mathbf{k}$ to $k_i = 2\pi\frac{n_i}{N}$, where $n_i < N$ is an
integer. For simplicity, we denote the vectors $\mathbf{k}$ on the
finite lattice as simply the integers $(n_1,n_2,n_3)$, and introduce
a factor of $2\pi/N$ in the exponential of Eq.~(\ref{eq:dischar}).
We then apply Eq.~(\ref{eq:proj}) using the characters given in
Eqs.~(\ref{eq:dischar}) and (\ref{eq:contchar}), modified for use on
the torus. For finite lattices, we find that the projection formula
reduces to the projection of the continuous rotation group to the
little group given by $\mathbf{k}$, independent of the lattice size.
We also find that representations labeled by different stars are
orthogonal. Therefore, the reduction of an arbitrary IR of
$\mathcal{G}$ labeled by $(\mathbf{k}, m_j)$ contains IRs of the
discrete group labeled by $\mathbf{k}$, and by $\alpha$ which
correspond to the reduction of $\mathrm{O}^\mathrm{D}$(2) to the
little group. These reductions for the various little groups of
$\mathrm{O}^\mathrm{D}_h$ and $m_j = 0, \frac{1}{2},...,3$ are given
in Tab.~\ref{tab:red}. As an example, we see that if
$\mathbf{k}=(1,0,0)$ and $m_j = 2$, then $E^{(1,0,0),2} =
\Gamma^{(1,0,0),B_1} \oplus \Gamma^{(1,0,0),B_2}$.

If $\mathbf{k}=0$, one reduces $\mathrm{O}^\mathrm{D}$(3) to
$\mathrm{O}^\mathrm{D}_h$. These results can be read off those given
by Johnson \cite{Johnson82} using the result that IRs of
$\mathrm{O}^\mathrm{D}$(3) with positive parity, $\pi = +1$,
correspond to the ``gerade" IRs (\textit{e.g.}\  $A_{1g}$) of
$\mathrm{O}^\mathrm{D}_h$ only, and those with $\pi =-1$ correspond
to the ``ungerade" IRs (\textit{e.g.}\  $A_{1u}$) only
\cite{Basak05,Morningstar99}.
\begin{table}
\caption{\label{tab:red}Reduction of the double cover of O(2) to the
possible little groups.  The double covers of the little groups
include the single valued IRs of the corresponding little groups in
$\mathrm{O}_h$ and double valued IRs corresponding to fermionic
continuum IRs.}
\begin{ruledtabular}
\begin{tabular}{lccccc}
  $m_j$ & $\mathrm{Dic}_4$ & $\mathrm{Dic}_3$ & $\mathrm{Dic}_2$ & $\mathrm{C_{4}}$ & $\mathrm{C_{2}}$\\
  \hline
  $0^+$         & $A_1$            & $A_1$            & $A_1$            & $A$ & $A$ \\
  $0^-$         & $A_2$            & $A_2$            & $A_2$            & $B$ & $A$ \\
  $\frac{1}{2}$ & $E_1$            & $E_1$            & $E$              & $E$ & $2B$ \\
  $1$           & $E_2$            & $E_2$            & $B_1 \oplus B_2$ & $A \oplus B$ & $2A$  \\
  $\frac{3}{2}$ & $E_3$            & $B_1 \oplus B_2$ & $E$              & $E$ & $2B$ \\
  $2$           & $B_1 \oplus B_2$ & $E_2$            & $A_1 \oplus A_2$ & $A \oplus B$ & $2A$ \\
  $\frac{5}{2}$ & $E_3$            & $E_1$            & $E$              & $E$ & $2B$ \\
  $3$           & $E_2$            & $A_1 \oplus A_2$ & $B_1 \oplus B_2$ & $A \oplus B$& $2A$  \\
  $\frac{7}{2}$ & $E_1$            & $E_1$            & $E$              & $E$ & $2B$ \\
  $4$           & $A_1 \oplus A_2$ & $E_2$            & $A_1 \oplus A_2$ & $A \oplus B$& $2A$\\
\end{tabular}
\end{ruledtabular}
\end{table}

\section{The moving harmonic oscillator}

We can now apply these results to a physical problem -- the
isotropic harmonic oscillator potential moving at some constant
(non-relativistic) velocity.  The solution for such a moving
potential is known in the continuum \cite{Duru89}, and the
wavefunctions are simply products of a translational piece and the
wavefunctions of the stationary potential.  Accordingly, the
energies just pick up an extra contribution due to the translational
energy of the center of mass, and the continuum spectrum is $E_{n,v}
= \frac{1}{2}mv^2 + E_n$ for translation at a constant velocity,
$v$, and the energies of the stationary oscillator, $E_n$.  In
Cartesian coordinates, the Hamiltonian for the stationary harmonic
oscillator potential is just the sum of three 1-D oscillators in
each of the coordinate directions. These coordinate pieces commute,
so the Schr\"{o}dinger equation is variable separable, the
wavefunctions are just products of the 1-D wavefunctions, and the
energies are just sums of the 1-D energies. Thus, the familiar
continuum levels are evenly spaced and have alternating parities
given as $(-1)^n$.

On an $N^3$ lattice, the Hamiltonian is still the sum of three
commuting pieces, where each piece is now an $N\times N$ matrix.
Using the finite difference approximation for the momentum, $p_x^2 =
-\hbar^2(\delta_{x+1,x'}+\delta_{x-1,x'}-2\delta_{x,x'})/a^2$, where
$a$ is the lattice spacing, then on a $3^3$ lattice, we find three
1-D levels with energies denoted as $E_0$, $E_1$, and $E_2$ in order
of increasing energy, and where $E_1 - E_0 > E_2 - E_1$. Unlike the
continuum, the energies are not evenly spaced, and they have
parities $+,+,-$ respectively. The energies of the 3-D oscillator on
the lattice are then given as $E_n = E_{n_x}+E_{n_y}+E_{n_z}$ for
$n_i=0,1,2$.
\begin{table}
\caption{(a)~The reduction of O(2) to the little group
$\mathrm{C_{4v}}$, which are obtained from Tab.~\ref{tab:red} by
taking the single valued representations of $\mathrm{Dic_4}$ on the
integer values of $m_j$. (b)~The reduction of the lattice oscillator
states to $\mathrm{C_{4v}}$ along with the parities of the states,
which correspond to the parities in the continuum limit.}
\begin{center}
\subtable[]{ \label{tab:shoa}
\begin{tabular}{lc}
\hline\hline
  $m_j$ & $\mathrm{C_{4v}}$ content \\
  \hline
  $0^+$ & $A_1$ \\
  $0^-$ & $A_2$ \\
  1 & $E$ \\
  2 & $B_1 \oplus B_2$ \\
  3 & $E$ \\
  4 & $A_1 \oplus A_2$ \\
\hline\hline
\end{tabular}
} \subtable[]{ \label{tab:shob}
\begin{tabular}{lll}
\hline\hline
  H.O. states & Parity & $\mathrm{C_{4v}}$ content \\
  \hline
  $|000\rangle$ & $+$ & $A_1$ \\
  $|100\rangle$ & $+$ & $2A_1 \oplus B_1$ \\
  $|200\rangle$ & $-$ & $A_1 \oplus E$ \\
  $|110\rangle$ & $+$ & $2A_1 \oplus B_1$ \\
  $|210\rangle$ & $-$ & $A_1 \oplus B_1 \oplus 2E$ \\
  $|220\rangle$ & $+$ & $B_2 \oplus E$ \\
  $|111\rangle$ & $+$ & $A_1$ \\
  $|211\rangle$ & $-$ & $A_1 \oplus E$ \\
  $|221\rangle$ & $+$ & $B_2 \oplus E$ \\
  $|222\rangle$ & $-$ & $B_2$ \\
\hline\hline
\end{tabular}
}
\end{center}
\end{table}

We now consider the inverse mapping of these lattice oscillator
states into continuum states of some angular momentum. This
inversion presents difficulty because as in Tab.~\ref{tab:red}, many
continuum states are mapped to each discrete state. For
$E_{|\mathbf{k}|,n}$ with $\mathbf{k} = 0$, this inverse mapping was
found by Johnson~\cite{Johnson82}.  We consider this mapping in the
case of a non-trivial momentum, $\mathbf{k} = (1,0,0)$ with little
group $\mathrm{C_{4v}}$ in $\mathrm{O}_h$. We see the degenerate
states are labeled by the vectors $\mathbf{k}'$ in the star of
$\mathbf{k}$ and the set of oscillator states $|n_xn_yn_z\rangle$
with energy $E_n$ for some fixed $n$. The characters of the
degenerate states as basis vectors for the representations of
$\mathcal{T}^3_{lat}\rtimes \mathrm{O}_h$ are then found, and the
reduction of the oscillator states to the IRs of the space group is
given in Tab.~\ref{tab:shob}. For simplicity, we omit the
translation labels in the table as the oscillator states trivially
contain only IRs labeled by $\mathbf{k} = (1,0,0)$.  Thus, the first
row of the table indicates that the moving oscillator state labeled
by $\mathbf{k} = (1,0,0)$ and $|n_xn_yn_z\rangle = |000\rangle$
contains the irreducible representation $\Gamma^{(1,0,0),A_1}$ of
$\mathcal{T}^3_{lat} \rtimes \mathrm{O}_h$. In the following, we
speak only of the rotational states, but the translational component
is implicit.

By comparison with the reduction of O(2) to $\mathrm{C_{4v}}$, as
given in Tab.~\ref{tab:shoa}, we see that the ground state
$|000\rangle$ is the lowest state containing $A_1$ and corresponds
to $m_j = 0^+$. The next state, $|100\rangle$ with $\mathrm{C_{4v}}$
content $2A_1 \oplus B_1$, has positive parity and corresponds to
the $m_j = 0^2,1,2$ doublet in the continuum (in the continuous
case, the degenerate states $|200\rangle_{cont}$ and
$|110\rangle_{cont}$ form a doublet with $j=0,2$.  For the
projection along some axis, this gives $m_j=0$ and $m_j=0,1,2$). The
states $|220\rangle$ and $|221\rangle$ both have positive parity and
the correct $\mathrm{C_{4v}}$ content to partner it, but the lower
energy state $|220\rangle$ is the correct partner. Without knowledge
of the continuum states, this assignment would be unclear.  The
state $|200\rangle$ has negative parity and is the lowest energy
$A_1 \oplus E$ state, and it corresponds to $m_j = 0,1$
(\textit{i.e.}\  $j = 1$). On this small lattice, the symmetric
boundary conditions have moved this level above part of the $j =
0,2$ doublet. The states $|110\rangle$ and $|210\rangle$ with
respective partners $|221\rangle$ and $|222\rangle$ are parts of the
continuum multiplets $j = 0,2,4$ and $j = 1,3$, but the multiplets
are not complete due to the small size of the lattice. Angular
momentum assignments for higher energy states are increasingly
difficult, and rely heavily upon knowledge of the continuum states.

The results found for the moving oscillator correspond to those
found in the case of $\mathbf{k}=0$ since the moving states of the
oscillator have the same angular momentum content as the states of
the stationary oscillator. However, states are now labeled only by
the projection of $j$ along $\mathbf{k}$.  Thus, states labeled by
$m_j = 0,1,...,j$ lie in different IRs and contribute to any state
with a given $j$.  With non-zero $\mathbf{k}$, the rotation group is
some subgroup of $\mathrm{O}_h$, and thus has fewer irreducible
representations. This leads to even more ambiguity in the inverse
mapping as the continuum states are mapped into fewer rotational
states on the lattice.

\section{Conclusion}
We have expanded upon the results for the mapping of angular
momentum states to the lattice and demonstrated the inverse mapping
for a system moving with non-zero linear momentum. Due to the
semidirect product structure of the groups we induce representations
of the groups from the subgroups of translations. Since arbitrarily
large translations are allowed, we consider the projection of states
on finite lattices and take the limit as the number of lattices
sites grows arbitrarily large. We find that the continuum states are
mapped to states with the same momentum vector, with the continuum
rotational states decomposing into the states in the little group of
the momentum vector.

In the example of the moving harmonic oscillator potential, we have
seen that, without additional information, the inverse mapping of
lattice states to continuum states is difficult for anything but the
lowest angular momentum states. In the continuum case these
ambiguities increase for non-zero momentum because continuum states
are mapped to fewer rotational states on the lattice. In addition,
states are now labeled by the projection of $j$ along the direction
of motion, so states with the same $j$ but different projections
$m_j$ belong to different irreducible representations.

\section{Acknowledgements}
DCM acknowledges support from the Yale College Dean's Research
Fellowship.  GTF would like to thank C.\ Morningstar, D.\ G.\
Richards and G.\ W.\ Kilcup for useful discussions in the early
phases of this work.

\bibliography{refs}
\bibliographystyle{apsrev}

\end{document}